\documentclass[aps,prl,showpacs,twocolumn,epsfig]{revtex4}
\usepackage[dvips]{graphicx}
\begin{document}

\draft
\title{Reaction cross section described by a black sphere 
approximation of nuclei}

\author{Akihisa Kohama,$^1$ Kei Iida,$^{1,2}$ 
and Kazuhiro Oyamatsu$^{1,3,4}$}
\affiliation{$^1$RIKEN 
(The Institute of Physical and Chemical Research),
2-1 Hirosawa, Wako-shi, Saitama 351-0198, Japan\\
$^2$RIKEN BNL Research Center, Brookhaven National Laboratory, 
Upton, New York 11973-5000\\
$^3$Department of Media Theories and Production, Aichi Shukutoku
University, Nagakute, Nagakute-cho, Aichi-gun, Aichi 480-1197, Japan\\
$^4$Department of Physics, Nagoya University, Furo-cho, Chigusa-ku, 
Nagoya, Aichi 464-8602, Japan}

\date{\today}

\begin{abstract}
     
     We identify a length scale that simultaneously accounts 
for the observed 
proton-nucleus reaction cross section and diffraction peak 
in the proton 
elastic differential cross section.  
This scale is the nuclear radius, $a$, 
deduced from proton elastic scattering data of incident 
energies higher than 
$\sim800$ MeV, by assuming that the target nucleus is 
a ``black'' sphere.  The 
values of $a$ are determined so as to reproduce the angle 
of the first 
diffraction maximum in the scattering data for stable nuclei.   
We find that the absorption cross section, $\pi a^2$, agrees 
with the empirical total reaction cross section for 
C, Sn, and Pb to within error bars.  This agreement persists 
in the case of 
the interaction cross section measured for a carbon target.
We also find that $\sqrt{3/5}a$ systematically 
deviates from the empirically deduced values 
of the root-mean-square matter radius
for nuclei having mass less than 
about 50, 
while it almost completely agrees with the deduced values for 
$A\gtrsim50$.  This tendency suggests a significant change of 
the nuclear 
matter distribution from a rectangular one for $A\lesssim50$, 
which is 
consistent with the behavior of the empirical charge distribution.

\end{abstract}


\maketitle

     The size of atomic nuclei is considered to be well deduced from
empirical data for the proton-nucleus elastic differential
cross section, $d\sigma_{\rm el}/d\Omega$, and the total reaction 
cross section, $\sigma_R \equiv \sigma_T-\sigma_{\rm el}$, 
where $\sigma_T$ is the total cross section.  So far, the analysis that 
respects both data in deducing the nuclear size has not been completed 
in particular for proton incident energies, $T_p$, higher than 800 MeV.  
Various approximate theories based on 
optical potentials have been proposed to reproduce the elastic 
scattering data, while they usually tend to overestimate the reaction 
cross section for 
$800$ MeV $\lesssim T_p \lesssim 1000$ MeV (e.g., Ref.\ \cite{Ray:PRC20} 
and references therein).

     In Ref.\ \cite{KIO}, we constructed a method for deducing the nuclear size
by focusing on the peak angle in the proton-nucleus elastic 
differential cross section measured at $T_p\gtrsim800$ MeV, where the 
corresponding optical potential is strongly absorptive.  In this method, we 
regard a nucleus as a ``black'' (i.e., purely absorptive) sphere of radius $a$,
and determine $a$ in such a way as to reproduce the angle of the observed 
first diffraction peak. If we multiply $a$ by $\sqrt{3/5}$, a ratio
between the root-mean-square and squared off radii for a rectangular 
distribution, the result for stable nuclei of $A\gtrsim50$ 
shows an excellent agreement with the root-mean-square radius, $r_m$,
of the matter density distribution as determined from conventional scattering 
theories so as to reproduce the overall diffraction pattern and analyzing power
in the proton elastic scattering. 

     In this paper, we extend such a previous analysis to the case of 
$A\lesssim50$, and find out a systematic deviation between $\sqrt{3/5}a$ 
and $r_m$.  We next show that the present method is effective at explaining 
the observed reaction cross sections for stable 
nuclei ranging from light to heavy 
ones.  In the black sphere approximation of a nucleus, where the 
geometrical cross section can be described by $\pi a^2$, $a$ plays the role of
a critical radius inside which the reaction with incident protons occurs.  
We find that $\pi a^2$ 
is consistent with the measured reaction cross section.
Consequently, the black sphere picture characterized by $a$ gives 
a unified basis for describing 
the reaction cross section and the elastic scattering data.
This simple formula for the reaction cross section, 
given by $\pi a^2$, does 
not include any adjustable parameter.  This is 
a feature more advantageous than the fitting formulas
proposed earlier \cite{Karol,Kox} on the basis of the $A^{1/3}$ law.

     We start with evaluations of $a$ for stable nuclei including
light elements by following a line of argument of Ref.\ \cite{KIO}.  
The center-of-mass (c.m.) scattering angle for proton elastic scattering
is generally given by $\theta_{\rm c.m.} = 2\sin^{-1}(q/2p)$ with
the momentum transfer, ${\bf q}$, and the proton incident momentum in the
c.m.\ frame, ${\bf p}$.  For the proton diffraction by a circular black
disk of radius $a$, we can calculate the value of $\theta_{\rm c.m.}$
at the first peak as a function of $a$.  (Here we define the zeroth
peak as that whose angle corresponds to $\theta_{\rm c.m.}=0$.)
We determine $a$ in such a way that this value of $\theta_{\rm c.m.}$
agrees with the first peak angle for the measured diffraction in
proton-nucleus elastic scattering, $\theta_M$. 
The radius, $a$, and the angle, $\theta_M$, are then related by
\begin{equation}
   2 p a \sin(\theta_M/2) = 5.1356 \cdots.
    \label{a}
\end{equation}
By setting
\begin{equation}
   r_{\rm BS}\equiv \sqrt{3/5}a,
   \label{rbb}
\end{equation}
we found \cite{KIO} that at $T_p\gtrsim800$ MeV,
$r_{\rm BS}$, estimated for heavy stable nuclei
of $A>50$, is within error bars consistent with 
the root-mean-square nuclear matter radius, $r_m$, deduced from 
elaborate analyses based on conventional scattering theory.  
Thus, expression (\ref{rbb}) works as a ``radius formula.'' 
The factor $\sqrt{3/5}$ comes from the assumption that the 
nucleon distribution is rectangular;
the root-mean-square radius of a rectangular distribution is a 
cutoff radius multiplied by $\sqrt{3/5}$.  Here we simply
extend this estimate of $r_{\rm BS}$ to lighter stable nuclei.
This extension of the black sphere analysis is reasonable as 
long as the
scattering is close to the limit of the geometrical optics. 
This condition is fairly well satisfied at least for 
$T_p\gtrsim800$ MeV, since $a/\lambda$, where $\lambda= 2\pi/p$ is 
the wave length of incident proton in the c.m.\ frame, 
is well above unity even for $^4$He.
As we shall see, the values of $r_{\rm BS}$ are systematically 
smaller than those of $r_m$ for $A<50$, whereas the values 
of $\pi a^2$
for C, Sn, and Pb agree well with the proton-nucleus reaction 
cross section data for $T_p\gtrsim800$ MeV.

\begin{figure}[t]
\begin{center}
\includegraphics[width=7cm]{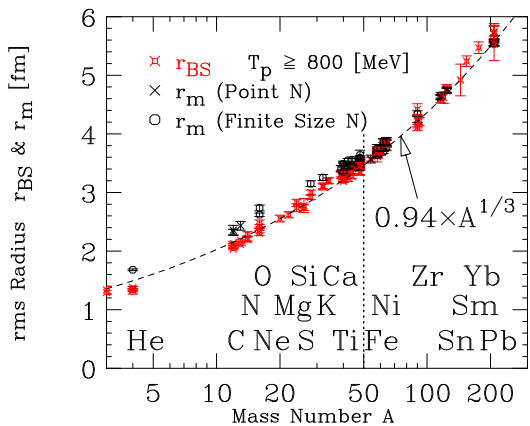}
\end{center}
\vspace{-0.5cm}
\caption{(Color) $r_{\rm BS}$ as a function of mass number, $A$.
For nuclei of $A>50$, we have plotted the values 
of $r_m$ adopted in Ref.\ \cite{KIO}. 
For the error bars in $r_{\rm BS}$ and $r_m$, see Ref.\ \cite{KIO}. 
For comparison we also plot the results for $r_m$ derived in the 
following references: 
For $^{12}$C, in Refs.\ \cite{Ray:PRC18a,Blan,Blan:PRC} ($T_p=800$ MeV); 
for $^{13}$C, in Refs.\ \cite{Ray:PRC18a,Blan:PRC} ($T_p=800$ MeV); 
for $^{16}$O, in Refs.\ \cite{Alk:NPA381,Chaum:PLB} ($T_p=1000$ MeV); 
for $^{28}$Si, in Ref.\ \cite{Alk:NPA381} ($T_p=1000$ MeV); 
for $^{32}$S, in Ref.\ \cite{Alk:NPA381} ($T_p=1000$ MeV); 
for $^{39}$K, in Ref.\ \cite{Alk:NPA381} ($T_p=1000$ MeV); 
for $^{40,48}$Ca, in Ref.\ \cite{Ray:PRC19} ($T_p=796$ MeV),
in Ref.\ \cite{Alk:NPA381} ($T_p=1000$ MeV); 
for $^{40,42,44,48}$Ca, 
in Ref.\ \cite{Igo:PLB} ($T_p=796$ MeV),
in Refs.\ \cite{Alk:NPA274,Chaum:PLB} ($T_p=1044$ MeV); 
for $^{46,48}$Ti, in Ref.\ \cite{Paul} ($T_p=799.3$ MeV); 
for $^{48}$Ti, in Ref.\ \cite{Alk:NPA274} ($T_p=1044$ MeV).
For $^4$He, the root-mean-square radius of the charge density distribution
deduced from electron elastic scattering data \cite{DeV:ATO,FB} is plotted 
instead of $r_m$, because no $r_m$ is available for this case. 
The crosses ($\times$) denote the root-mean-square matter radii of the point 
nucleon distributions, and the circles ($\circ$) denote those folded with 
the nucleon form factor.
The dashed curve represents $0.94A^{1/3}$ fm. 
The dotted line shows $A=50$.
} 
\end{figure}

     The values of $r_{\rm BS}$, which are obtained for stable nuclei ranging 
from He to Pb from the proton elastic scattering data for $T_p\gtrsim800$ MeV,
are plotted in Fig.\ 1, together with the values of $r_m$ estimated in 
earlier investigations.  
In collecting the data, we have made access to
Experimental Nuclear Reaction Data File (EXFOR) \cite{exfor}. 
In deriving $r_{\rm BS}$ for $^3$He, 
we have used the data for $T_p=800$ MeV \cite{Geso} 
and for $T_p=1040$ MeV \cite{Frasca}. 
Since the peak position is not clear, 
we do not include the data for $T_p=1000$ MeV \cite{Alk:PLB85} in this analysis. 
For $^{4}$He, we have used the data for $T_p=1000$ MeV \cite{Pale,Alk:JETPL26},  
$T_p=1050$ MeV \cite{Baker}, and $T_p=1150$ MeV \cite{Asla}.  
We remark that, for $T_p=890,991$ MeV, no peak has been identified 
since the measured cross sections are limited to very forward 
direction \cite{Greben}. 
For C and heavier isotopes, we have used the following data:
$^{12}$C ($T_p=800$ MeV) \cite{Blan,Blan:PRC,Blan:PRC23};
$^{13}$C ($T_p=800$ MeV) \cite{Blan,Blan:PRC,Blan:PRC32};
$^{12}$C ($T_p=1000$ MeV) \cite{Alk:PL42B};
$^{12}$C ($T_p=1040$ MeV) \cite{Bertini};
$^{14}$N ($T_p=800$ MeV) \cite{Blan:PRC25a,Blan:PRC32};
$^{14}$N ($T_p=1000$ MeV) \cite{Alk:pre}; 
$^{16}$O ($T_p=800$ MeV) \cite{Adams,Blesz}; 
$^{16}$O ($T_p=1000$ MeV) \cite{Alk:pre,Alk:JETPL28};
$^{20}$Ne ($T_p=800$ MeV) \cite{Blan:PRC30};
$^{22}$Ne ($T_p=800$ MeV) \cite{Blan:PRC38};
$^{24}$Mg ($T_p=800$ MeV) \cite{Blan:PRC20,Blan:PRC25,Blan:PRC37};
$^{26}$Mg ($T_p=800$ MeV) \cite{Blan:PRC25,Blan:PRC37};
$^{28}$Si ($T_p=1000$ MeV) \cite{Alk:NPA381,Alk:SovJ}; 
$^{32,34}$S ($T_p=1000$ MeV) \cite{Alk:pre,Alk:SovJ}; 
$^{39}$K ($T_p=1000$ MeV) \cite{Alk:pre,Alk:JETPL18};
$^{40,42,44,48}$Ca ($T_p=796$ MeV) \cite{Igo:PLB};
$^{40}$Ca ($T_p=797.5$ MeV) \cite{Blesz:PRC25}
$^{40}$Ca ($T_p=800$ MeV) \cite{Blesz,Ray:PRC23};
$^{40}$Ca ($T_p=1000$ MeV) \cite{Alk:pre,Alk:JETPL18};
$^{40}$Ca ($T_p=1044$ MeV) \cite{Alk:NPA274};
$^{42,44}$Ca ($T_p=800$ MeV) \cite{Ray:PRC23};
$^{42,44}$Ca ($T_p=1044$ MeV) \cite{Alk:NPA274};
$^{48}$Ca ($T_p=800$ MeV) \cite{Ray:PRC23};
$^{48}$Ca ($T_p=1000$ MeV) \cite{Alk:pre};
$^{48}$Ca ($T_p=1044$ MeV) \cite{Alk:NPA274};
$^{46,48}$Ti ($T_p=799.3$ MeV) \cite{Paul};
$^{48}$Ti ($T_p=1044$ MeV) \cite{Alk:NPA274}. 
We do not include the data 
for $^{9}$Be ($T_p=1000$ MeV) \cite{Zhusu}, 
$^{12}$C ($T_p=1000$ MeV) \cite{Pale}, 
nor $^{16}$O ($T_p=1000$ MeV) \cite{Pale}
since the peak positions are not clear. 

   It is interesting to note that $r_{\rm BS}$ agrees almost completely
with $r_m$ for $A>50$, as shown in Ref.\ \cite{KIO}, while it is 
smaller than $r_m$ by about 0.2 fm for $A<50$.  
In order to clarify this deviation, 
we exhibit the difference, $r_{\rm BS}-r_m$, in Fig.\ 2. 
The drastic change in the difference around $A\sim50$ implies a possible 
change in the form of the real nucleon distribution; the rectangular 
distribution as assumed in the present black sphere model may well 
simulate the real distribution at $A>50$, while for $A<50$ the
real distribution is quite different from the rectangular one in such 
a way that the portion of the real distribution farther than $a$
is relatively large.  This feature is suggested by the empirical 
charge distribution deduced from the electron-nucleus elastic scattering 
\cite{DeV:ATO}, which shows a Gaussian-like form rather than a rectangular 
one for light nuclei.

     This feature of the nucleon distribution is expected 
to be reflected
by size-sensitive observables for which empirical data are
available for stable nuclei ranging from light to heavy ones.
Such observables include $1s$ states of pionic atoms and
isoscalar giant resonance energies; the isoscalar part of
the pion-nucleus optical potential \cite{Fried}
and the inertia associated with the resonances \cite{BG} 
are related to the nucleon distribution.

\begin{figure}[t]
\begin{center}
\includegraphics[width=7cm]{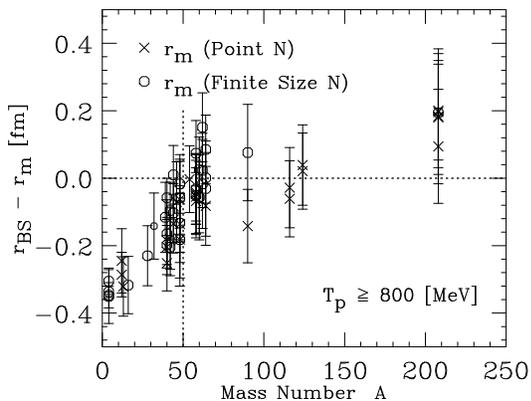}
\end{center}
\vspace{-0.5cm}
\caption{The difference, $r_{\rm BS} - r_m$,
as a function of mass number, $A$.
The crosses ($\times$) and the circles ($\circ$) are calculated 
from the corresponding values of $r_m$ in Fig.\ 1.  
The dotted line shows $r_{\rm BS}=r_m$ and $A=50$.
} 
\end{figure}

     We can see from Fig.\ 1 that $r_{\rm BS}$, 
if having its staggering 
smoothed out, behaves as $\sim0.94A^{1/3}$ fm.  
This suggests that $r_{\rm BS}$ and hence $a$ provides 
a measure of the 
reaction cross section, since the empirical values are 
known to behave 
roughly as $\propto A^{2/3}$ \cite{Bauho}.  It is thus interesting to 
compare the black sphere cross section with the empirical data. 

     We proceed to calculate the proton-nucleus total
reaction cross section from the present black sphere model.  
Our model regards it as the geometrical cross section, 
\begin{equation}
  \sigma_{\rm BS}\equiv\pi a^2.
   \label{rrf}
\end{equation}
Here we assume that the incident protons are point particles, 
and that the 
incident protons, if touching the target point nucleon distribution, 
contribute to the reaction cross section by yielding excitations 
associated with internucleon motion.  
Our model thus predicts vanishing 
cross section for the proton-proton case.  This is reasonable since 
the proton-nucleon reaction cross section is relatively small 
for $T_p\lesssim1$ GeV \cite{PDG,Arndt}.

     By substituting the values of $a$ determined by Eq.\ (\ref{a}) 
into Eq.\ (\ref{rrf}), we evaluate $\sigma_{\rm BS}$ 
for stable nuclei.
In Fig.\ 3, the results are plotted together with the empirical
data on $\sigma_R$ available for 800 MeV $<T_p < 1000$ MeV 
\cite{Bauho}.  We can see an excellent agreement between 
$\sigma_{\rm BS}$ and the empirical values although the comparison 
is possible only for C, Sn, and Pb.  
For these nuclei, as shown in Table I, $\sigma_{\rm BS}$ 
and $\sigma_R$ agree with each other within the error bars.
Note that we do not use any adjustable parameter here.
We thus see the role played by 
$\sigma_{\rm BS}$ in predicting $\sigma_R$, and this is useful
for nuclides for which elastic scattering data are available but 
no data for $\sigma_R$ are available.  
On the other hand, if $\sigma_R$ is measured,
one can deduce $a$ from Eq.\ (\ref{rrf}).  In this case, $a$ can be
regarded as a ``reaction radius,'' inside which the reaction 
with incident protons occurs.
In a real nucleus, this radius corresponds to a radius at which 
the mean-free path of incident protons is of the order 
of the length of the penetration.

\begin{figure}[t]
\begin{center}
\includegraphics[width=7cm]{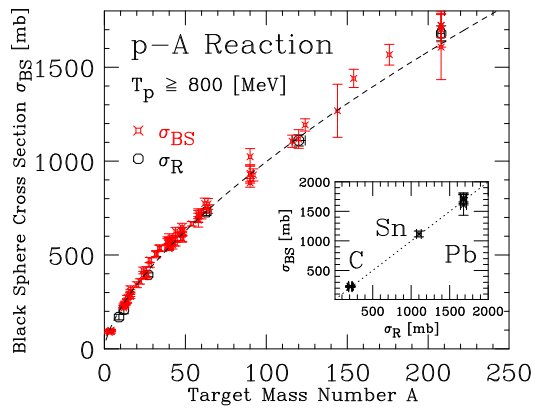}
\end{center}
\vspace{-0.5cm}
\caption{(Color)
The absorption cross section, $\sigma_{\rm BS}$, of a proton of 
$T_p\gtrsim800$ MeV by a target nucleus of mass number $A$. 
For comparison, we plot the empirical data for the proton-nucleus
total reaction cross section, $\sigma_R$ ($\circ$), 
which are listed in Ref.\ \cite{Bauho}
for $^9$Be, $^{27}$Al, C, Cu, Sn, and Pb.
For the latter four elements the value of $\sigma_R$
is the average over the isotopic abundance in a target.
For these data, we set $A$ as the mass number of the most 
abundant isotope  
and assign the uncertainty in $A$ due to the natural abundance. 
The dashed curve denotes $(5/3)\pi(0.94A^{1/3})^2$ fm$^2$.
Inset: $\sigma_{\rm BS}$ vs \ $\sigma_R$ for C, Sn, and Pb.
The dotted line represents $\sigma_{\rm BS}=\sigma_R$.
} 
\end{figure}

\begin{table}[b]
\caption{Values of $\sigma_R$ and $\sigma_{\rm BS}$ 
for C, Sn, and Pb. 
The values of $\sigma_R$ \cite{Bauho} are for natural targets, 
while those of $\sigma_{\rm BS}$ are for 
$^{12}$C, $^{120}$Sn, and $^{208}$Pb, 
the elastic scattering data for which are taken from
the references shown in the last column.
}
\begin{center}
\begin{tabular}{|c||c|cc|}   
\hline
Target &  $\sigma_R$~[mb]       ~($T_p$~[MeV]) 
       &  $\sigma_{\rm BS}$~[mb]~($T_p$~[MeV]) & Ref.  \\
\hline
C      &  209     $\pm$  22     ~(860) 
       &  217.003 $\pm$   3.45  ~(800)         & \cite{Blan} \\
       &                              
       &  227.788 $\pm$   3.46  ~(800)         & \cite{Blan:PRC23}\\
       &                         
       &  228.233 $\pm$   9.20  ~(1000)        & \cite{Alk:PL42B} \\
       &                          
       &  237.562 $\pm$   6.69  ~(1040)        & \cite{Bertini} \\
\hline 
Sn     & 1100     $\pm$  30     ~(860) 
       & 1117.486 $\pm$  49.6   ~(800)         & \cite{Gazza} \\
\hline
Pb     & 1680     $\pm$  40     ~(860)  
       & 1723.590 $\pm$  88.5   ~(800)         & \cite{Blan} \\
       &                       
       & 1721.680 $\pm$  63.6   ~(800)         & \cite{Hof:PRL} \\
       &                       
       & 1606.919 $\pm$ 173.0   ~(800)         & \cite{Gazza} \\
       &                          
       & 1701.161 $\pm$  94.1   ~(1000)        & \cite{Alk:pre} \\
\hline
\end{tabular}
\end{center}
\label{table1}
\end{table}

    It is natural to generalize the definition of $\sigma_{\rm BS}$
given by Eq.\ (\ref{rrf}) to the case of the nucleus-nucleus reaction 
cross section.  We simply set 
\begin{equation}
   \sigma_{\rm BS} = \pi(a_1 + a_2)^2, 
   \label{sigbb}
\end{equation}
where $a_1 (a_2)$ is the black sphere radius of a projectile
(target) nucleus, which we determine from the proton elastic 
differential cross section data for $T_p\gtrsim800$ MeV. 

      In the case of nucleus-nucleus collisions, rather than 
to measure the 
reaction cross section $\sigma_R$, it is more convenient 
to measure the 
interaction cross section, 
$\sigma_I \equiv \sigma_R - \sigma_{\rm inela}$, 
where $\sigma_{\rm inela}$ is the cross section for 
inelastic channels.
It is interesting to compare
the measured values of $\sigma_I$ with the corresponding values of
$\sigma_{\rm BS}$ given by Eq.\ (\ref{sigbb}).  The results are given
in Fig.\ 4, together with the empirical data on $\sigma_I$ 
measured with a $^{12}$C target for stable nuclei of 
$E/A \gtrsim 800$ MeV \cite{Ozawa}. 

    We find from Fig.\ 4 that $\sigma_{\rm BS}$ agrees 
within error bars with the empirical values 
of $\sigma_I$ except for a few cases. 
This result reassures the role played by the black sphere radius as a 
reaction radius.  We also note the tendency that $\sigma_{\rm BS}$ is 
larger than $\sigma_I$.  This is consistent with the facts that 
$\sigma_R>\sigma_I$ and that, for a $^{12}$C projectile,  
$\sigma_{\rm BS}$ is much closer to the empirical value 
of $\sigma_R$ \cite{Jaros} than that of $\sigma_I$.

\begin{figure}[t]
\begin{center}
\includegraphics[width=7cm]{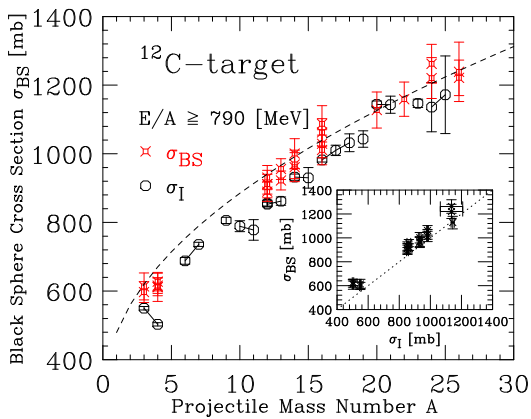}
\end{center}
\vspace{-0.5cm}
\caption{(Color online) The absorption cross section 
$\sigma_{\rm BS}$ 
for a projectile stable 
nucleus of $A<30$ and a $^{12}$C target.  
We fix $r_{\rm BS}=2.086\pm0.05$ fm 
for $^{12}$C.  For comparison, we plot the interaction cross section, 
$\sigma_I$ ($\circ$), measured at $E/A\gtrsim800$ MeV
for a projectile of $^{3, 4}$He, $^{6, 7}$Li, $^9$Be, $^{10, 11}$B, 
$^{12, 13}$C, $^{14, 15}$N, $^{16, 17, 18}$O, 
$^{19}$F, $^{20, 21}$Ne, $^{23}$Na, and $^{24, 25}$Mg
incident on a $^{12}$C target \cite{Ozawa}.
The dashed curve represents $(5/3)\pi(2.086+0.94A^{1/3})^2$ fm$^2$. 
Inset: $\sigma_{\rm BS}$ vs \ $\sigma_I$ for
$^4$He, $^{12, 13}$C, $^{14}$N, $^{16}$O, $^{20}$Ne, and $^{24}$Mg. 
The dotted line represents $\sigma_{\rm BS}=\sigma_I$.
}
\end{figure}

      It is also interesting to evaluate $\sigma_{\rm BS}$ for highly
neutron-enriched nuclei once the proton elastic differential cross 
section is measured at high incident energies.  If the proton 
reaction 
cross section $\sigma_R$ or the interaction cross 
section $\sigma_I$ is measured simultaneously, 
one could compare the result 
with $\sigma_{\rm BS}$.
If $\sigma_{\rm BS}$ is significantly smaller than $\sigma_{R~(I)}$,
it would suggest that the significant contribution to $\sigma_{R~(I)}$
comes from the tail region farther than $a$. 
This could imply the possible presence of a neutron halo \cite{Ozawa}.
The comparison between $\sigma_{\rm BS}$ and $\sigma_{R~(I)}$
provides an important check of the widely accepted speculation that 
one can deduce the halo structure from measurements of
the interaction cross section \cite{IKO}.

     In summary we have extended our previous analysis of the nuclear 
size for $A>50$ \cite{KIO}, which is based on the black sphere 
assumption of a nucleus, to lighter stable nuclei.  
The black sphere radius, $a$, has been determined so as to 
reproduce the angle of the first diffraction maximum in the proton
elastic scattering data.  We have thus finalized a systematic 
analysis of the existing data for proton elastic scattering off 
stable nuclei ranging from He to Pb at proton incident energies above 
$\sim800$ MeV.  We have two significant results.  First, 
the values of $r_{\rm BS}$ obtained 
from Eq.\ (\ref{rbb}) are systematically 
larger than the root-mean-square radius $r_m$ deduced 
from elaborate scattering 
theory for $A<50$, although they agree quite well with each other
for $A>50$.  This suggests a significant deviation of the nucleon
distribution from the rectangular one for $A<50$.  Second, the 
absorption cross section, $\pi a^2$, is consistent with 
the empirical total reaction cross section for C, Sn, and Pb.
This consistency persists in the case of the interaction cross
section measured for a carbon target.  We thus see the dual 
role of $a$ as a black sphere radius and as a reaction radius.

   Elastic scattering and total reaction cross section data for
heavy unstable nuclei are expected to be provided by radioactive
ion beam facilities, such as GSI, Darmstadt, and the Radioactive
Ion Beam Factory in RIKEN.  Experiments at RIKEN are planned for
a beam energy of a few hundreds MeV per nucleon.  It is thus
important to study the energy dependence of the black sphere radius,
$a$.  As a first trial, we have studied the case of $^{208}$Pb, for
which not only the de Broglie wavelength of the incident proton
but also the mean-free path of the proton in the nuclear interior
is sufficiently short compared to the nuclear radius, and confirmed
that the resultant $\sigma_{\rm BS}$ agrees with the measured value
of $\sigma_R$ within the error bars for the proton incident energy
down to a few hundreds MeV.  The energy dependence is found to be
similar to that of the nucleon-nucleon total cross section.  This
is consistent with the fact that the black sphere radius corresponds
to the reaction radius.  We remark that the energy dependence of $a$
may well modify the relation between $a$ and $r_m$ 
\cite{KIO2}.

     Towards future possible application of the black sphere
model to neutron-rich unstable nuclei, we here give a couple of
examples, $^{6, 8}$He and $^{11}$Li, for which empirical
information on the proton-nucleus total reaction cross section
is available at relatively high energy.  Hereafter we will
estimate the black sphere radius $a$ from $\sigma_R=\pi a^2$
in the absence of the data for the first peak in the elastic
diffraction pattern.  From the empirical data,
$\sigma_R=161.3 \pm 3.7$ mb for $^6$He ($E/A=721$ MeV) and
$\sigma_R=197.8 \pm 3.5$ mb for $^8$He ($E/A=678$ MeV),
measured by Neumaier {\it et al}. \cite{Neu:NPA}, we estimate
$a=2.27 \pm 0.03$ fm and $r_{\rm BS}=1.76 \pm 0.03$ fm for $^6$He, and
$a=2.51 \pm 0.02$ fm and $r_{\rm BS}=1.94 \pm 0.02$ fm for $^8$He.
The values of $r_m$ deduced from the elastic scattering
data \cite{Alk:NPA712} are 
$r_m=2.45 \pm 0.10$ fm for $^6$He and
$r_m=2.53 \pm 0.08$ fm for $^8$He.  
We thus see a difference
between $r_m$ and $r_{\rm BS}$ of order 0.7 fm, which is
significantly larger than that for light stable nuclei (see Fig. 2.)
For $^{11}$Li, we take note of the data for the interaction cross
section obtained for a proton target: $\sigma_I=276 \pm 8$ mb at
800 MeV$/A$ \cite{Taniha}.  
If we assume $\sigma_R = \sigma_I$, we obtain
$a = 2.96 \pm 0.04$ fm and $r_{\rm BS} = 2.30 \pm 0.04$ fm.
Since the typical value of $r_m$ amounts to about 3.1 fm 
\cite{Ozawa},
we obtain $r_m - r_{\rm BS}$ of about 0.8 fm, which is similar to
the case of neutron-rich He isotopes.  This leads to an interesting
implication that such a difference is typical of nuclei having a
neutron halo.  In order to justify this implication, however, it is
inevitable to confirm, by future experiments, the assumption that
even for neutron-rich unstable nuclei, $\sigma_R$ agrees with the
black sphere cross section determined from the first peak angle in
proton-nucleus elastic scattering differential cross section.

     We acknowledge K. Yazaki for his invaluable suggestions 
and comments. 
We also acknowledge the members of Japan Charged-Particle 
Nuclear Reaction 
Data Group (JCPRG), especially N. Otuka, 
for kindly helping us collect the various data sets. 
K.I. thanks the Institute for Nuclear Theory at the
University of Washington for its hospitality and the
Department of Energy for partial support during the completion 
of this work.
K.O. thanks Aichi Shukutoku University for its domestic research
program, and Department of Physics at Nagoya University for 
its hospitality.

\end{document}